\documentclass[11pt,a4paper]{article} 
\usepackage{times}

\usepackage[utf8]{inputenc} 

\usepackage{geometry} 

\usepackage{graphicx} 

\usepackage[utf8]{inputenc} 
\usepackage{etex}
\usepackage{graphicx}
\usepackage{dcolumn}
\usepackage{amssymb,amsthm,amscd,amsbsy,array}
\usepackage{amstext}
\usepackage{amsmath,dsfont}
\usepackage{cancel}
\usepackage{soul} 
\usepackage{graphics,xcolor}
\usepackage{bm}
\usepackage{booktabs} 
\usepackage{array} 
\usepackage{paralist} 
\usepackage{verbatim} 
\usepackage{subfig} 
\usepackage{authblk} 
\usepackage{bbding} 

\usepackage{multirow}

\usepackage[numbers,sort&compress]{natbib}

\numberwithin{equation}{section}

\let\ssection=\section
\renewcommand{\section}{\setcounter{equation}{0}\ssection}
\textheight=
230mm 
\newcommand{\half}{{\frac{1}{2}}}
\def\2{{\half}}

\def\beq{\begin{equation}}
\def\eeq{\end{equation}}
\def\beqa{\begin{eqnarray}}
\def\eeqa{\end{eqnarray}}

\def\barray{\left(\begin{array}}
\def\earray{\end{array}\right)}
\def\barraynb{\begin{array}}
\def\earraynb{\end{array}}

\def\smallover#1/#2{\hbox{$\textstyle\frac{#1}{#2}$}} %

\newcommand{\Tr}{\mathrm{Tr}}

\DeclareMathOperator{\sech}{sech}
\DeclareMathOperator{\csch}{csch}

\usepackage[colorlinks=true, pdfstartview=FitV, linkcolor=blue, citecolor=blue, urlcolor=blue]{hyperref} 
\usepackage{color}

\newcommand{\gb}{\colorbox{green}}

\def\benu{\begin{enumerate}}
\def\eenu{\end{enumerate}}

\def\?{{\;\gb{\,?\,} \;}}

\usepackage{fancyhdr} 
\usepackage{sectsty}
\allsectionsfont{\sffamily\mdseries\upshape} 

\usepackage[nottoc,notlof,notlot]{tocbibind} 
\usepackage[titles,subfigure]{tocloft} 

\title{Displacement Memory Effect from Supersymmetry}    
\author[]{Erdal Catak\thanks{ecatak@istanbul.edu.tr} }
\affil[]{\vspace{0.2cm} \small Department of Physics, Istanbul University, 34134 Istanbul, T\"urkiye}

\author[]{Mahmut Elbistan\thanks{mahmut.elbistan@bilgi.edu.tr } }
\affil[]{\vspace{0.2cm}\small Department of Energy Systems Engineering, Istanbul Bilgi University, Istanbul 34060, T\"urkiye}

\author[]{Mustafa Mullahasanoglu\thanks{mustafa.mullahasanoglu@std.bogazici.edu.tr } }
\affil[]{\vspace{0.2cm}\small Department of Physics, Bogazici University, 34342 Bebek, Istanbul, T\"urkiye
}
\affil[]{\vspace{0.2cm}\small
Feza Gursey Center for Physics and Mathematics, Bogazici University, 34684, Kandilli,
Istanbul, T\"urkiye}

\date{} 

\begin{document}

\maketitle


\begin{abstract}

We explain the recent results on the displacement memory effect (DME) of plane gravitational waves using supersymmetric quantum mechanics. This novel approach stems from that both the geodesic and the Schrödinger equations are Sturm-Liouville boundary value problems. Supersymmetry provides a unified framework for the Pöschl-Teller and the Scarf profiles and yields the critical values of the associated wave amplitudes for DME in a natural way. Within our framework, we obtain a compact formula for DME in terms of the asymptotic values of the superpotential and the geodesics. In addition, this new technique enables us to build plane and gravitational waves with 2-transverse directions using superpartner potentials. Lastly, we study DME within a singular wave profile inspired by supersymmetric quantum mechanics, which demonstrates the broader applicability of our method.
\vspace{0.5cm}


\end{abstract}

\pagebreak

\tableofcontents

\newpage

\section{Introduction}

The memory effect of plane gravitational waves can be divided into two types: \emph{the displacement memory effect (DME)} \cite{ZeldovichPolnarev} and \emph{the velocity memory effect (VME)} \cite{EhlersKundt, Sou73, GriPol}. These effects are of interest as they can hopefully be captured by  future gravitational wave detectors \cite{Sun:2022pvh, Agazie:2025oug}. In this paper, we focus our attention at the problem of DME of the plane gravitational waves (GWs) with a novel perspective by applying techniques borrowed from supersymmetric quantum mechanics (SUSY QM)  \cite{Gendenshtein:1983skv,Dutt:1986va,Cooper:2001zd, Khare:2004kn, Correa:2012np}.

Previous studies (see \cite{Zhang:2024uxf, Elbistan:2023qbp} and references therein) suggested that  plane waves might only have a VME: \emph{after the passing of a short burst of a gravitational wave, transverse geodesics have a non-vanishing constant velocity}. Only very recently it has been reported that plane GWs, under special conditions, can also exhibit a DME, that is, a \emph{permanent displacement with vanishing velocity after the wave has passed}. According to \cite{Zhang:2024uyp, Zhang:2024tey, Zhang:2025lxs} (see also \cite{BenAchour:2024ucn}), DME occurs only in particular setups with fine-tuned parameters. Our main motivation is to investigate these recent results about the DME for the Pöschl-Teller (PT) (\ref{PTprof}) and the Scarf GW (\ref{scarfpro}) profiles from the point of view of SUSY QM. We show that \emph{there is a hidden supersymmetry behind the DME phenomena of the PT and the Scarf GWs which unifies those profiles and determines the critical wave amplitudes in an elegant manner.} 

Our starting point is that the geodesics with DME conditions satisfy a Sturm-Liouville (SL) problem with Neumann boundary conditions. We are already familiar with such equations from another context, namely non-relativistic QM. 
Therefore, the DME conditions of the plane GWs can be sought for by using the Schrödinger theory. The desired geodesics could, in principle, be found analytically by the methods of SUSY QM \cite{Gendenshtein:1983skv,Dutt:1986va,Cooper:2001zd, Khare:2004kn}. 

This common mathematical origin of geodesics of plane waves and quantum mechanics sheds light on the recent results of those \emph{half-waves} and \emph{magical numbers} of the former \cite{Zhang:2024uyp, Zhang:2024tey, Zhang:2025lxs} by relating them to the bound states and the quantum numbers of the latter. Namely, we begin with a superpotential (\ref{superscarf}) and create partner potentials (\ref{scarfpart}) including both the Pöschl-Teller and the Scarf profiles at the same time. Then, SUSY leads to those particular setups and solutions in a natural way. 
We also borrow a singular potential from SUSY QM and discuss the possibility of finding DME within singular GWs.

Plane waves, being models for gravitational radiation, are already significant spacetimes on their own \cite{phonebook}. They are Penrose limits of any spacetime \cite{Penrose}, and lower dimensional mechanics can be embedded into them via the Eisenhart-Duval lift \cite{Eisenhart, DuvalBargmann, DGH}. 

We show, for the first time we believe,  that the necessary conditions for DME relate plane waves to integrable potentials in SUSY QM.  Using this novel interrelation, we obtain the geodesics with DME by using SUSY methods in a compact and beautiful manner. SUSY also helps us to understand the conditions to build exact GWs exhibiting DME in both directions.

Let us note that the geodesic equations of spinning particles in gravitational backgrounds were worked out using supersymmetry in  \cite{Gibbons:1993ap}. Our study differs from that both in its content and in its scope.

The outline of the paper is the following: In Section \ref{sectionpwave}, we introduce our main motivation, which is the DME of plane waves. Then, in Section \ref{sectionwavesusy}, we apply SUSY methods of quantum mechanics to plane wave geodesics and discuss how DME occurs.  4-dimensional plane waves and their DME are discussed in Section \ref{4dsusygw}. In Section \ref{Sectsingular}, we comment on the DME for singular wave profiles. In the last section, Section \ref{disout}, we summarize our results, add our comments, and ask further questions.


\section{Plane waves and memory effects}
\label{sectionpwave}

In standard Brinkmann coordinates, a generic $d+2$-dimensional plane wave metric  is given as 
\begin{equation}
\label{planew}
ds^2 = \delta_{ij}dX^i dX^j+ 2 dU dV + K_{ij}(U) X^i X^j dU^2,  
\end{equation}
where $U, V$ are light-cone coordinates and $X^i$ with $i =1,..., d$, are transverse coordinates. The profile $K_{ij}(U)$ is a symmetric $d\times d$ matrix. The plane wave (\ref{planew}) is known to be an exact plane GW if its profile is traceless.

To discuss DME and VME, we consider the geodesic motion. Plane waves (\ref{planew})  have a covariantly constant null Killing vector $\frac{\partial}{\partial V}$. Thus, the coordinate $U$ can be chosen as an affine parameter, and  geodesic equations for the remaining coordinates turn out to be
\begin{subequations}
\label{geoplane}
\begin{align}
&\frac{d^2 X^i}{dU^2} - K^i{}_{j}X^j =0, \\
&\frac{d^2 V}{dU^2} + \frac{1}{2} \frac{d K_{ij}}{dU} X^i X^j + 2 K_{ij} X^i \frac{dX^j}{dU} =0.
\end{align}
\end{subequations} 
We can immediately solve the $V$-equation (\ref{geoplane}b) if the transverse geodesics $X^i(U)$  are known, 
\begin{equation}
\label{vsln}
V(U) = - \frac{1}{2} X^i \frac{d X_i}{dU} + c_1 U + c_2.
\end{equation}
The constants $c_1$ and $c_2$ are to be determined from the initial conditions. 
So, the most crucial part of the geodesic problem (thus, of the memory effect) is to solve (\ref{geoplane}a), which is an SL problem when $K_{ij}$ depends on $U$. 

Only a few exact solutions are known, and we usually confine ourselves to numerical methods (see, e.g, \cite{Elbistan:2023qbp}). Analytical or numerical, both types of studies show that the geodesics of plane GWs have, generically, a non-vanishing constant velocity after the wave has passed, confirming the VME  considered in \cite{EhlersKundt, Sou73, GriPol}. 
Therefore, it comes as a surprise that the Gaussian profile with $d=1$ 
\begin{equation}
\label{Gausspro}
 K^G (U) = \frac{k}{\sqrt{\pi}} e^{-U^2}, 
\end{equation}
yields  DME if certain critical values are chosen for the amplitude, $k=k_{crit}$ \cite{Zhang:2024uyp, Zhang:2024tey}. This confirms the statement of Zel'dovich and Polnarev \cite{ZeldovichPolnarev}, who advocated that the relative velocity will be zero and there will only be pure displacement at distant times. 
In that manner, the geodesic should satisfy the Neumann boundary conditions  
\begin{equation}
\label{Neumann}
\frac{d X^i}{dU}|_{U\to \pm\infty} =0,
\end{equation}
and the related equation (\ref{geoplane}a) becomes an SL boundary value problem. 

In their quest for such magical values $k=k_{crit}$, authors of  \cite{Zhang:2024uyp, Zhang:2024tey} were motivated to replace the Gaussian profile (\ref{Gausspro}) with that of the Pöschl-Teller (PT)
\begin{equation}
\label{PTprof}
K^{PT}(U)= -\frac{k}{4\cosh^2U}.   
\end{equation}
This profile yields a good approximation of the Gaussian (\ref{Gausspro}).
While for $K^{G}(U)$ only numerical solutions are available, the geodesics of the PT profile (\ref{PTprof}) with (\ref{Neumann}) can be found analytically: DME occurs for $k_{crit}=4m(m+1)$ where $m$ is a positive integer. 

Moreover, a very recent paper \cite{Zhang:2025lxs} reports that the Scarf profile
\begin{equation}
\label{scarfpro}
K^{scarf}(U)= -2g \frac{\sinh U}{\cosh^2U},
\end{equation} 
can be used as an approximation to the derivative of (\ref{Gausspro}) proposed to describe flyby \cite{Gibbons:1971wsk}.
Akin to the PT profile (\ref{PTprof}), the Scarf profile is also capable of having a DME if magic parameter values are chosen. It is even possible to create an exact plane GW with $d=2$, with both components yielding DME geodesics. Similar results were found numerically for the derivative Gaussian profile  \cite{Zhang:2025lxs}. 

Let us mention that a PT-type wave profile (\ref{PTprof}) was also employed in \cite{Chakraborty:2022hrp, Bhattacharya:2025ljc} to study DME and VME within wormhole geometries. In \cite{Chakraborty:2020uui}, the same profile was used in Kundt wave spacetimes to indicate a relation between the curvature and the memory. The authors showed the existence of DME numerically.
In \cite{Chakraborty:2019yxn}, similar profiles are adopted to study the B-memory of the vacuum plane waves and in \cite{Bhattacharjee:2020vfb} DME was studied in asymptotic black hole spacetimes.

Leaving those numerical studies aside, we employ SUSY to obtain exact  results both for the PT profile (\ref{PTprof}) and for the Scarf profile (\ref{scarfpro}). The list of exactly solvable potentials in \cite{Dutt:1986va,Cooper:2001zd, Khare:2004kn} includes them within a single superpotential.
Thus, the methods of SUSY are directly applicable to their geodesic problem which can be treated as a non-relativistic Schrödinger equation. Using this relation, we reveal the origin of DME of plane waves in the context of SUSY QM, which restores the magical values of parameters and necessary conditions naturally.
We believe that this interrelation of the DME problem of plane GWs with SUSY QM is interesting in its own right.

\section{Creating plane and gravitational waves via SUSY QM}
\label{sectionwavesusy}

The transverse part of the geodesic equations in plane waves (\ref{geoplane}a), together with (\ref{Neumann}), constitute an  SL boundary value problem\footnote{For simplicity, the profile $K_{ij}(U)$ is taken to be diagonal.}. Therefore, analytically solvable geodesic equations of this form are expected to correspond to exact solutions derived with elegant methods in SUSY QM.

Here, we will explicitly reveal the correspondence between geodesics of plane waves in Section \ref{sectionpwave} and quantum mechanics by deriving the results of  \cite{Zhang:2024uyp, Zhang:2024tey, Zhang:2025lxs} with the methods of SUSY. 

Our approach has a couple of advantages: 
\begin{enumerate}
\item Using a superpotential $W(U)$ (\ref{ladder}) allows us to unify the recent works about DME of plane GWs in an elegant way.

\item Incorporation of supersymmetric methods explain the magical amplitude values and the half waves of DME for those particular GWs. 

\item We can construct further wave profiles with solvable geodesics that can mimic more realistic but not analytically solvable ones.
\end{enumerate}

\subsection{SUSY QM: A brief review}
\label{sectionsusy}
 
Below, we provide a brief,  self-contained summary of SUSY methods. The interested reader may consult \cite{Dutt:1986va,Cooper:2001zd, Khare:2004kn} for more details.
 
We begin with recalling the main elements of SUSY QM, namely its superpotential $W(U)$ and ladder operators $L,  L^\dagger$,
\begin{equation}
\label{ladder}
L = \frac{d}{dU} + W(U), \quad L^\dagger = - \frac{d}{dU} + W(U). 
\end{equation}
They factorize the Hamiltonians $H_1$ and $H_2$ as
\begin{subequations}
\label{partners}
\begin{align}
H_1 &= L^\dagger L = -\frac{d^2}{dU^2} + V_1, \quad \quad  V_1 = W^2 - \frac{d W}{dU}, \\
H_ 2 &= L L^\dagger = -\frac{d^2}{dU^2} + V_2, \quad  \quad  V_2 = W^2 + \frac{d W}{dU}.
\end{align}
\end{subequations}
$V_1$ and $V_2$ (\ref{partners}) are partner potentials sharing the same spectrum, except for the zero-energy ground state of $V_1$. 
The time-independent Schrödinger equation for the first potential $V_1$ is 
\begin{equation}
\label{Scheqn}
 \left(-\frac{d^2}{dU^2} + V_1\right)\psi_n = E_n  \psi_n,
\end{equation}
where $n=0, 1,2, ...$ indexes the bound state. Up to a normalization factor, the zero energy ground state $\psi_0(U)$ can be found as
\begin{equation} 
\label{ground}
\psi_0 (U)= e^{- \int^U  du\ W(u)}, \quad E_0 =0.
\end{equation}

If $V_1$ and $V_2$ are \emph{shape invariant} \cite{Dutt:1986va} i.e.,  
\begin{equation}
\label{shapein}
V_2(U; A_1) = V_1(U; A_2) + f(A_1), 
\end{equation}
where $A_{1,2}$ are parameters\footnote{Recall that plane wave profiles (\ref{Gausspro}), (\ref{PTprof}) and (\ref{scarfpro}) also have parameters like $k, g,..$.}, energy eigenvalues and eigenstates of (\ref{partners}) can be computed easily. Both $f(A_1)$ and $A_2(A_1)$ are functions of the first parameter $A_1$. This crucial property of shape invariance of SUSY QM will allow us to derive DME of plane GWs.

Due to (\ref{shapein}), discrete eigenstates are found one by one 
\begin{equation}
\label{shapepsi}
\psi_n(U; A_1) = L^\dagger(U; A_1)\psi_{n-1}(U; A_2). 
\end{equation}
For instance, if $A_1$ and $A_2$ are related by a simple relation $A_1 -A_2=1$, then the excited state wave functions can be computed with the subsequent action of $L^\dagger$ (\ref{ladder})
\begin{equation}
\label{shapegeo}
\psi_n (U; A) = L^\dagger(U; A) L^\dagger(U; A-1)...L^\dagger(U; A-n+1)\psi_0(U; A-n),
\end{equation}
where we set $A_1 =A$. In that case, (\ref{shapepsi}) simplifies as
\begin{equation}
\label{shapepsis}
\psi_n(U; A) =  \left(   - \frac{d}{dU} + W(U; A) \right)\psi_{n-1}(U; A-1). 
\end{equation}
For now, the parameter $A$ can take any value. However, when this framework is applied to the geodesics, the DME problem puts restrictions on such parameters.


\subsection{Unified wave profiles via SUSY QM}

Consider the following superpotential
\begin{equation}
\label{superscarf}
W(U)= A\, \tanh U + B\, \sech U,
\end{equation}
where $A, B$ are free parameters.
It figures in the list of shape invariant potentials in  \cite{Gendenshtein:1983skv, Dutt:1986va, Cooper:2001zd, Khare:2004kn}\footnote{For a historical remark, see \cite{Natanson}.}. Out of it, one can build two partner potentials (\ref{partners})
\begin{subequations}
\label{scarfpart}
\begin{align}
V_1 & = (B^2 - A^2 -A)\ \sech^2 U + B(2A +1)\ \sech U \tanh U + A^2,       \qquad \qquad\\
V_2 &=(B^2 - A^2 +A)\ \sech^2 U + B(2A -1)\ \sech U \tanh U + A^2.       \qquad \qquad 
\end{align}
\end{subequations}
$V_1$ and $V_2$ are related via (\ref{shapein}) where $A$ is the shape invariance parameter
\begin{equation}
\label{shapeinc}
V_2(U; A) = V_1(U; A-1) + 2A-1.
\end{equation}
Using (\ref{shapeinc}) and shifting $A$, the exact bound state spectrum for $V_1$ (hence for $V_2$) can be found as \cite{Cooper:2001zd}
\begin{equation}
 E_{n} = A^2 - (A- n)^2, \quad n=0, 1,\cdots A.
\end{equation}
The ground state with $ E_{0} =0$ corresponds to $n=0$, $n=1$ is the first excited bound state, etc. Note that $V_{1,2}$ vanish as $U \to \pm \infty$. Below, we will show that they admit wave functions with vanishing derivatives at infinities so that they satisfy the DME conditions (\ref{Neumann}).

We cast the associated Schrödinger equation (\ref{Scheqn}) into the form of a geodesic equation (\ref{geoplane}a)
\begin{equation}
\label{Schscarf1}
 \frac{d^2 \psi_n}{dU^2}  -\left[ (A-n)^2 +  (B^2 - A^2 -A)\ \sech^2 U + B(2A +1)\ \sech U \tanh U  \right] \psi_n = 0.
\end{equation}
Up to a normalization factor, the ground state wave function can be obtained from (\ref{ground}),
\begin{equation}
\psi_0 (U)= e^{-\left( A\ln(\cosh U ) + B \tan^{-1} (\sinh U ) \right)}.
\end{equation}

In order to pass to the geodesics (\ref{geoplane}), we make the replacement $\psi(U) \to X(U)$  in the above expressions. It will be immediately recognized that  (\ref{Schscarf1}) is composed of 2 parts: the PT profile (\ref{PTprof}) in \cite{Zhang:2024uyp, Zhang:2024tey} and the Scarf profile (\ref{scarfpro}) in \cite{Zhang:2025lxs}. Therefore, we  achieve our first goal of a unified framework. \emph{This unified view will allow us to understand DME of the related profiles in a systematic way by explaining the quantized values of the magical amplitudes}.

\subsection{DME of plane GWs}
\label{DMEsusytech}
In this section, we build the related geodesics using SUSY methods and investigate their DME.  Our calculations are made for $d=1$ case. However,  they are valid for $d=2$ plane and GWs of Section \ref{4dsusygw} as their profiles $K_{ij}(U)$ are diagonal. 

Let us begin with $n=1$. Using the ladder operator $L^\dagger$ (\ref{shapegeo}), we get 
\begin{eqnarray}
\label{X1m}
X_1(U) &=& L^\dagger(U; A) X_0(U; A-1), \nonumber \\
&=& \sech^{A-1}U\ e^{-B \tan^{-1}(\sinh U)} \left[ (2A-1)\tanh U + 2B \sech U \right].
\end{eqnarray}
Recall that $A$ is the shape invariance parameter, thus it cannot be equal to 0. The negative values of $A$ yield diverging geodesics with diverging velocities at infinity. Thus, $X_1(U)$ satisfies the DME condition (\ref{Neumann}) provided that $A\geq 1$
\footnote{It is a half wave in the language of  \cite{Zhang:2024uyp, Zhang:2024tey,Zhang:2025lxs}.}. One can also see that as $U \to \pm\infty$, $X_1(U) \neq 0$ only if  $A=n=1$.

Subsequent SUSY Darboux transformations (\ref{shapegeo}) generate other geodesics with higher \emph{half-wave numbers} $n$. 

With this unified approach, we will show below that the PT and the Scarf profiles in \cite{Zhang:2024uyp, Zhang:2024tey, Zhang:2025lxs} are actually \emph{two different faces of the same theory}, viewed from the perspective of SUSY QM.  Both profiles are rooted in (\ref{superscarf}), hence derivable from  (\ref{scarfpart}).  

For this, an important point is the value of the free parameter $A \geq n$. If we set $A =n$, the formula for geodesics (\ref{shapepsis}) simplifies in terms of $W(U)$ as
\begin{equation}
\label{Xnn}
X_n (U; A=n) =  W(U; n) X_{n-1}(U; n-1) - \frac{d}{dU} X_{n-1}(U; n-1).
\end{equation}
The DME is the difference between the asymptotic values of geodesics:
\begin{equation}
\label{DME}
\Delta X_n \equiv X_n(U \to\infty; A=n) - X_n(U\to-\infty; A=n). 
\end{equation}
Substituting (\ref{Xnn}) into (\ref{DME}), we obtain
\begin{eqnarray}
\Delta X_n = W(\infty; n) X_{n-1}(\infty; n-1) - W(-\infty; n) X_{n-1}(-\infty; n-1),
\label{DMESUSY}
\end{eqnarray}
as our geodesics satisfy the Neumann boundary condition (\ref{Neumann}). Therefore, SUSY QM yields a compact and useful formula  (\ref{DMESUSY}) for the DME of plane waves in terms of the asymptotic values of superpotential and the preceding geodesics\footnote{Our formula (\ref{DMESUSY}) might be somewhat reminiscent of Christodoulou's nonlinear memory \cite{Christodoulou:1991cr}.}. 
Via (\ref{DME}) we observe that a non-vanishing DME requires parity odd geodesics and our second formula (\ref{DMESUSY}) relates this condition to the parity of the superpotential.  
But the individual parts of $W(U)$ (\ref{superscarf})  have different parity properties. Thus, we investigate each part separately and show that these magical numbers come out of our unified picture in a beautiful manner.

\subsubsection{The Pöschl-Teller profile}
\label{PT1d}

To recover the PT-type profile (\ref{PTprof}), we need to set 
\begin{subequations}
\label{ptset}
\begin{align}
B(2A+1) &= 0, \\
A-n&=0,
\end{align}
\end{subequations}
in (\ref{Schscarf1}). While the first condition above eliminates the parity-odd Scarf term, the second one selects the \emph{highest excited state by quantizing the parameter $A$}. An integer $A$ leads to the so-called reflectionless PT problem. In that case, the geodesic equation turns out to be
\begin{equation}
\label{PT1}
\frac{d^2 X_n}{dU^2}  +  n(n+1) \text{sech}^2(U)  X_n = 0,
\end{equation}
cf. (\ref{geoplane}a).
As we are dealing with a shape invariant profile, the bounded geodesics can be found via (\ref{shapegeo}), where we set $A=m$ first,
\begin{equation}
\label{shapeingeo}
X_n (U; m) = L^\dagger(U, m)\  L^\dagger (U, m-1) ... L^\dagger(U, m+1-n)   X_0 (U, m-n),
\end{equation}
where $ X_0 (U, m) = \sech^m U$ and $m=1,2,3,\cdots$.

Observe that the seed geodesic $X_0 (U, m)$ is parity even. However, when $B=0$, the superpotential (\ref{superscarf}) becomes parity odd. In addition, each Darboux transformation $L^\dagger$ (\ref{shapeingeo}) changes the parity. Therefore, $\Delta X_n$ (\ref{DMESUSY}) vanishes for even $n$.

For example, if we take $n=1$, we obtain 
\begin{eqnarray}
\label{x1um}
X_1(U; m) &=& L^\dagger(U, m) X_0(U, m-1), \nonumber \\
&=& \left(  -\frac{d}{dU} + m \tanh U\right) \sech^{m-1} U = (2m-1)\sech^{m-1} U \tanh U. 
\end{eqnarray}
The Neumann boundary condition (\ref{Neumann}) 
holds for any $m \geq 1$.  Further setting $m=n=1$, we get 
\begin{equation}
\label{x1u}
X_1 = \tanh U,
\end{equation}
which is consistent with \cite{Zhang:2024uyp, Zhang:2024tey}. The related DME can be computed easily from (\ref{DMESUSY}) as
\begin{equation}
\Delta X_1 = 2.
\end{equation}
So, there is a net displacement, as expected from the parity.

Similarly,  $n=2$ geodesic is obtained using (\ref{shapeingeo}) as
\begin{equation}
\label{x2um}
X_2(U; m) = (2m-3)\Big(2 (m-1) \sech^{m-2}U  + (1-2m)\sech^mU\Big),
\end{equation}
and for any $m\geq 2$, it satisfies the DME condition (\ref{Neumann}). When $m=2$, the geodesic becomes 
\begin{equation}
\label{x2u}
X_2 = 3\tanh^2U -1, 
\end{equation}
with $\Delta X_2 =0$ from both (\ref{DME}) and (\ref{DMESUSY}). 

For $n=3$, the solution of the geodesic equation (\ref{PT1}) is
\begin{equation}
\label{x3um}
X_3 (U; m) = (2m-5)(2m-3)\Big( 2(m-2)\sech^{m-3} U + (1-2m) \sech^{m-1}U \Big)\tanh U .
\end{equation}
As before, non-vanishing displacement occurs only when $m=n=3$, such that 
\begin{equation} 
\label{x3u}
X_3(U; m=3) = 6 \tanh U - 15 \tanh U \sech^2U,  \quad  \Delta X_3 (m=3) = 12.
\end{equation}

With SUSY, one can go on and build geodesics for any $n$ and  we observe that $\Delta X \neq 0$ only if $m=n$ and 
$n$ is an odd integer. We conclude that \emph{the geodesics endowed with DME are nothing but the highest excited states of the analogous quantum mechanical problem with Neumann boundary condition. }

\vspace{0.2cm}

\textbf{Longitudinal motion and memory:}

To compute the $V$-component of the geodesics for the first two values of $n$, we substitute our solutions $X(U)$ i.e., (\ref{x1u}) and (\ref{x2u})  into (\ref{vsln}), respectively. We observe that part of the longitudinal motion organizes itself as another particular solution of the PT problem. For instance, 
\begin{eqnarray}
V_1 (U) &=& -\frac{1}{2} \tanh U\sech^2U + c_1U + c_2 = -\frac{1}{10} X_1(U; m=3)+ c_1U + c_2  , \\
V_2(U)&=& -3 \tanh U \sech^2U (2-3\sech^2U) + c_3 U +c_4 = -\frac{1}{35} X_3(U; m=5) +c_3 U +c_4, \qquad \quad
\end{eqnarray}
where $c_1, .., c_4$ are constants. $X_1(U; m=3)$ and $ X_3(U; m=5)$ can be found from (\ref{x1um}) and (\ref{x3um}), respectively. These motions exhibit VME when the initial constants $c_1, c_3$ do not vanish.

\subsubsection{The Scarf profile}
\label{scarfprofile}

In order to obtain the geodesic equation for the Scarf profile (\ref{scarfpro}), we need to satisfy another set of equations
\begin{subequations}
\label{scarfset}
\begin{align}
A-n &=0, \\
B^2 -A^2 -A &=0,
\end{align}
\end{subequations}
in (\ref{Schscarf1}). 
While the first equation quantizes the parameter $A$, (\ref{scarfset}b) solves for $B$ non-trivially as
\begin{equation}
\label{droot}
B = \pm\sqrt{n(n+1)},  \quad  A=n.
\end{equation}
where $n=0,1,2,\cdots$. \emph{Pay attention at  that $B$ has double roots} as in \cite{Zhang:2025lxs}.  This multiplicity will allow us to extend our arguments to exact plane gravitational waves with 2-transverse directions\footnote{That extension was first done in \cite{Zhang:2025lxs}. Here, we elaborate on their results using SUSY methods.} in Section \ref{4dsusygw}.

When $d=1$ (\ref{Schscarf1}) reduces to a geodesic equation (\ref{geoplane}) 
\begin{equation}
\label{Schscarf2}
\frac{d^2 X_n}{dU^2} \mp (2n+1) \sqrt{n(n+1)}\ \text{sech}\, U \ \text{tanh}\, U\ X_n =0,
\end{equation}
augmented with the boundary condition (\ref{Neumann}).
Thus, \emph{supersymmetry leads to a quick and elegant derivation of the Scarf geodesic equation which mimics the derived Gaussian flyby profile}. In \cite{Zhang:2025lxs} the above equation was obtained through the hypergeometric equation and solved with the Nikiforov-Uvarov method. SUSY provides a more economical way to find these solutions.

From the viewpoint of quantum mechanics, $X_n(U)$  is not the zero-energy ground state of the Scarf potential (\ref{scarfpro}). Rather, it corresponds to a particular parametrization of its parameters $A, B$ with the quantum number $n$ (\ref{droot}). Let us derive DME within this profile. 

The seed geodesic can be found via (\ref{ground}) as
\begin{equation}
X_0(U) = \sech^mU\ e^{-B\tan^{-1}(\sinh U)},
\end{equation}
where $A=m$ and $B =  \pm\sqrt{m(m+1)}$. As before, $m$ is a positive integer. When $n=1$, a family of geodesics labeled by $m$ can be obtained from (\ref{shapegeo}) as
\begin{equation}
\label{X1mscarf}
X_1(U; m) =\left[ (2m-1)\tanh U + 2B \sech U \right]  \sech^{m-1}U\ e^{-B \tan^{-1}(\sinh U)} .
\end{equation}
 (\ref{X1mscarf}) yields a non-vanishing DME only when $m=n=1$. In this case, it reduces to 
\begin{equation}
\label{X1mnscarf}
X_1(U; m=1) = \left[\tanh U + 2B \sech U \right]  e^{-B \tan^{-1}(\sinh U)}. 
\end{equation}
Let us emphasize that (\ref{X1mnscarf}) satisfies (\ref{Schscarf2}) with $n=1$. The related DME can be computed either directly from the geodesic solution above or from (\ref{DMESUSY}) 
\begin{eqnarray}
\Delta X_1 = 2\cosh\left(\frac{B\pi}{2}\right) = 2\cosh\left(\frac{\pi}{ \sqrt{2}}\right),  \quad    B =\pm \sqrt{2},
\end{eqnarray}
which is independent of the sign of $B$.

Applying a second Darboux transformation, we generate $X_2(U; m)$. In order to satisfy (\ref{Schscarf2}), we set $m=2$ and get 
\begin{equation}
X_2(U; m=2) = \left[2+ 21\sech^2U+ 8B\tanh U \sech U \right]\ e^{-B \tan^{-1}(\sinh U)}, 
\end{equation}
with $B^2=6$. The DME of this second geodesic is sensitive to the sign of $B$ 
\begin{equation}
\Delta X_2 = -4 \sinh\left( \frac{B\pi}{2} \right),
\end{equation}
as $B=\pm \sqrt{6}$.   

By virtue of the supersymmetric transformations (\ref{Xnn}), one may obtain $n \geq 3$ geodesics with non-zero DME in a straightforward manner. 

We would like to emphasize the non-trivial quantization of $B$ (\ref{droot}) in the Scarf profile. This sign ambiguity makes it totally different from the PT case where the sole parameter $A$ takes only positive integer values.  Another difference is that the Scarf profile may yield a non-vanishing DME even for $m\neq n$. The same results can be obtained via the second potential $V_2$ (\ref{scarfpart}b) by shifting $A \to A+1$.

\vspace{0.2cm}

\textbf{Longitudinal motion and memory:}

In order to illustrate the longitudinal memory for the Scarf profile, we construct $V_1(U)$  (\ref{vsln}) as
\begin{eqnarray}
V_1(U) &=& - \frac{1}{2} X_1(U; m=1) \frac{d X_1(U; m=1)}{dU} + c_1 U +c_2, \\
&=& - 3 \sech U \Big( 5\tanh U \sech U + B(1+ \sech^2 U)\Big) e^{-2B\tan^{-1}(\sinh U)}  + c_1 U +c_2. \qquad  
\end{eqnarray}
Since
$
\frac{dV_1}{d U}|_{U \to \pm \infty} = c_1,$
the coordinate $V$ exhibits a VME unless $c_1$ vanishes. This result can be extended to any $V_n(U)$ with $n \geq 2$.

\section{4-dimensional plane waves}
\label{4dsusygw}

Those $d=1$ cases studied above can be extended to plane waves with 2-transverse directions. For this, we choose a diagonal profile $K_{ij}(U)$ in (\ref{planew}) and set $X^i =\{X, Y\} $.

\subsection{Exact plane gravitational wave with the Scarf profile}

We begin with the Scarf case. As in  \cite{Zhang:2025lxs}, the double root property of $B$ (\ref{droot}) allows us to construct an exact plane GW (vacuum solution of Einstein's equations) mimicking flyby. The associated profile is diagonal and traceless
\begin{equation}
\label{Kscarf}
K_{scarf}
= \begin{pmatrix}
(2n+1)|B| \sech U \tanh U  &  0 \\
0  & -(2n+1)|B| \sech U \tanh U\,
\end{pmatrix},
\end{equation}
as $B = \pm |B|$ and $|B| =\sqrt{ n(n+1)}$ with $n$ being a positive integer. As we discussed before, the 
$n=1$ and $n=2$ geodesics with their DMEs can be simply extracted from Section \ref{scarfprofile} as
\begin{subequations}
\begin{align}
X_1(U) &=  \left[\tanh U + 2\sqrt{2} \sech U \right]  e^{-\sqrt{2} \tan^{-1}(\sinh U)},   \quad \quad    \Delta X_1 = 2\cosh\left( \frac{\pi}{\sqrt{2}} \right), 
\\[6pt]
Y_1(U) &=  \left[\tanh U - 2\sqrt{2} \sech U \right]  e^{\sqrt{2} \tan^{-1}(\sinh U)},   \qquad \quad    \Delta Y_1 = 2\cosh\left( \frac{\pi}{\sqrt{2}} \right),
\end{align}
\end{subequations}
and 
\begin{subequations}
\begin{align}
X_2 &=  \left[2+ 21\sech^2U+ 8\sqrt{6}\tanh U \sech U \right] e^{-\sqrt{6} \tan^{-1}(\sinh U)}, \quad \Delta X_2 = -4\sinh\Big( \frac{\sqrt{6}\pi}{2}\Big), 
\\[6pt]
Y_2&=   \left[2+ 21\sech^2U- 8\sqrt{6}\tanh U \sech U \right] e^{\sqrt{6} \tan^{-1}(\sinh U)}, \qquad \Delta Y_2 = 4\sinh\Big( \frac{\sqrt{6}\pi}{2}\Big).
\end{align}
\end{subequations}
These results are consistent with  \cite{Zhang:2025lxs}.

It is also possible to find lightcone geodesic $V_n(U)$ by combining solutions of those transverse geodesics above. As explained in Section \ref{scarfprofile}, the longitudinal memory will be a VME. 

We may construct further plane wave solutions where $m > n$, see (\ref{X1mscarf}). In that case, the trace of the profile is non-zero, i.e., $\Tr{K} = 2(m-n)^2$ (\ref{Schscarf1}). One can get a non-vanishing DME when, for instance, $n=2$ and $m=3$.

We emphasize that the construction of the exact plane GW (\ref{Kscarf}) with $\Tr(K)=0$ (\ref{Kscarf}) depends heavily on the non-trivial quantization of the parameter $B$ with sign ambiguity. Based on this, one can combine the terms with different signs. This is no longer true for the PT plane waves as it will be outlined below.

\subsection{Pöschl-Teller plane waves}
\label{PTplane4d}

In \cite{Zhang:2024tey}, it was argued that when an exact plane GW is built out of a PT profile, DME can be obtained only in one direction. The other direction should be put to zero via appropriate initial conditions. SUSY provides an argument for this: Unlike as in the Scarf case, the parameter $A$ of the PT profile has a single root (\ref{ptset}b) which can take only positive integer values. The shape invariance condition allows us to build a family of profiles labeled by $n$ but always with the same sign. 

To understand this better, we construct a 4-dimensional plane wave out of the supersymmetric partners (\ref{scarfpart}). The result will be a null-fluid spacetime, $\Tr(K)\neq 0$, with the following diagonal profile 
\begin{equation}
\label{KPT}
K_{PT}
= \begin{pmatrix}
(m-n)^2 - m(m+1)\sech^2U  &  0 \\
0  &  (m-n)^2 - m(m-1)\sech^2U\,
\end{pmatrix},
\end{equation}
where parameters $m$ and $n$ are integers. From Section \ref{PT1d}, we recall that DME occurs only when $m=n$. That simplifies (\ref{KPT}) and puts it into its final form. Then, the resulting geodesic equations (\ref{geoplane}a) are
\begin{subequations}
\label{KPT2}
\begin{align}
\left(\frac{d^2}{dU^2} + m(m+1)\sech^2U\right) X&=0, \\
\left(\frac{d^2}{dU^2} + m(m-1)\sech^2U\right) Y&=0.
\end{align}
\end{subequations}
Due to shape invariance, the solutions  $X(U)$ and $Y(U)$ are interrelated: they match with the exception for the ground state of $X$. If one sets $m=1$, the solution for the first one is $X= \tanh U$ (\ref{x1u}). The second equation, on the other hand,  becomes that of a free particle and confirms the claim made in  \cite{Zhang:2024tey}. Recall that choosing $m$ as an integer leads to a \emph{reflectionless PT potential}~: the reflection coefficient  vanishes just like for a free particle. The shape invariance integrability condition (\ref{shapeinc}) puts a limitation on the parameter $A$ so that it takes only positive values. Therefore, PT plane wave profiles (\ref{KPT}) and (\ref{KPT2}) always share the same sign and it is impossible to build an exact plane GW in that case.

Other choices of $m > 1$ yield DME again in one direction. For instance, when $m=2$, $X = 3\tanh^2U -1$ and $Y= \tanh U$. Therefore,  using  superpartners in the profile yields a tower of geodesics exhibiting DME in alternating directions.

\section{Modeling singular profiles with SUSY}
\label{Sectsingular}

Our final goal is to show that SUSY provides further insight and predicts DME for other GWs.

In \cite{Andrzejewski:2018pwq, Andrzejewski:2018zby}, memory effects of the profile $K(U)=U^{-4}$ were discussed. Later, the authors in \cite{Zhang:2024uxf} obtained a  DME within the same singular plane GW. 

There exist also exact plane GWs with a  $U^{-2}$ singularity near the origin. The simplest example is the one with  profile 
\begin{equation}
K(U)=\frac{k}{U^2}\rm{diag}(1, -1),
\end{equation}
 whose symmetries were studied in \cite{Zhang:2019gdm}. The Lukash GW \cite{Elbistan:2020dxz, Zhang:2021lrw}, which is endowed with an extra symmetry suffers from the same singularity\footnote{That $\frac{1}{U^2}$ singularity appears in nuclear physics too, see \cite{Elbistan:2018rds} and references therein.}. 

Now, we show that SUSY QM allows us to search for DME also within such singular plane waves. The related profile can be constructed using the following superpotential
\begin{equation}
\label{singW}
W(U) = A \coth U,
\end{equation}
which is singular at $U=0$. In this case the partner potentials become 
\begin{subequations}
\label{singV}
\begin{align}
V_1 &= A^2 + A(A+1)\csch^2U, \\
V_2&= A^2 + A(A-1)\csch^2U.
\end{align}
\end{subequations}
They are shape invariant with $A \to A-1$, thus the system is integrable.  Note that the superpotential (\ref{singW}) is parity odd; e.g.,  $W(\infty)= -W(-\infty)=1$. 
In order to obtain a non-vanishing DME, we select $A=n$, where $n$ labels the geodesics. Then, the associated geodesic equation becomes
\begin{equation}
\left(\frac{d^2}{dU^2} - n(n+1)\csch^2U\right)X=0,
\end{equation}
and its solutions can be immediately found with the technique summarized in Section \ref{DMEsusytech}. For instance,
\begin{subequations}
\begin{align} 
X_1&= \coth U, \qquad \qquad \ \ \Delta X_1= 2, \\
X_2&= 3\coth^2U-1, \qquad \Delta X_2=0.
\end{align}
\end{subequations}
These solutions blow up at $U=0$, as expected.
Using the parity argument given in (\ref{DME}) and (\ref{DMESUSY}), we deduce that only geodesics labeled with odd $n$ have a non-vanishing DME. For instance, we obtain $\Delta X_3= 12$ without even solving for $X_3(U)$. 

If we were to build a plane wave including the partner potential $V_2$, it would be pretty much similar to the PT plane wave described in Section (\ref{PTplane4d}). On the other hand, if we build an exact plane GW with 
\begin{equation}
\label{KsingW}
K_{sing}
= \begin{pmatrix}
 n(n+1)\csch^2U  &  0 \\
0  &  - n(n+1)\csch^2U\,
\end{pmatrix},
\end{equation}
we obtain DME only in the $X$-sector with the same argument as in Section \ref{PTplane4d}.

Based on our model (\ref{singV}), we conclude that DME can also be found in GWs  with $U^{-2}$ singularity. Of course, the memory effects in singular profiles should be handled with more care. Our arguments here can be taken as a necessary first step towards a complete solution which will be reported elsewhere.


\section{Discussion}
\label{disout}

In this work, we reveal a hidden supersymmetry in the DME problem of particular plane waves and recover the recent results of \cite{Zhang:2024uyp, Zhang:2024tey, Zhang:2025lxs} within the supersymmetric framework. However, our study goes beyond a mere rederivation in several ways. First of all, SUSY methods unify the Scarf and Pöschl-Teller GW profiles in an elegant way, showing that they are two different reductions of a common ancestor (\ref{superscarf}). The reduction process amounts to satisfying a couple of equations either (\ref{ptset}) or (\ref{scarfset}), which automatically yields the amplitudes of the wave profiles: no fine-tuning is necessary.  For PT profile, this reduction quantizes the shape invariance parameter $A$ and sets $B=0$. Therefore, we are left with a family of profiles labeled by $n\geq 1$. Due to the parity arguments, geodesics only with odd $n$ have a non-vanishing DME. When it comes to the Scarf case, we observe that the parameter $B$ is quantized non-trivially with a sign ambiguity (\ref{droot}). As a result, one may build a $4$-dimensional exact plane GW (\ref{Kscarf}) using this property which is absent in the PT case. Thanks to shape invariance, 
the complicated solutions for the geodesics can be found readily using Darboux transformations. See Table \ref{table} for some of our results.

\begin{table}[htbp]
\caption{SUSY generated DME}
\centering
\begin{tabular}{c c c c c }
\hline\hline
W(U) & Profile & Quantized parameters & DME ($\Delta X_1$) & Parity  \\ [0.5ex]
\hline 
$A\tanh U +B\sech U$ & Pöschl-Teller & $A=n,\ B=0$ & $2$ & even \\
$A\tanh U +B\sech U$ & Scarf & $A=n, \ B= \pm\sqrt{n(n+1)}$ & $2 \cosh(\frac{\pi}{\sqrt{2}})$ & odd \\
$A\coth U -B\csch U$ & Singular & $A=n, \ B=0$ & $2$ &  even  \\  [1ex]
\hline  
\end{tabular}
\label{table}
\end{table}

In addition, we propose a hypothetical 2-d transverse PT plane wave, which explains its half DME. $4$-dimensional Scarf exact GW fits to our framework because of the sign ambiguity in its quantized parameter $B$  (\ref{droot}).

Lastly, we study DME in GWs with $U^{-2}$ singularity. Our SUSY motivated model (\ref{singV}) predicts the existence of the DME for such type of GWs. Because of the singularity, this investigation requires much care, and we reserve its complete treatment for a follow-up study where we also plan to work out DME in cosmological Lukash GW.

Another research problem is to look for DME in the GW originating from the Kepler potential \cite{Zhang:2019gtt}. Thanks to SUSY, we may determine the solutions of the related SL problem that leads us to  BJR coordinates. Based on this, we can compare the Carroll symmetry generators of plane GWs with the symmetries of the associated quantum mechanical problem.  

There remain further questions to be explored. The list of SUSY QM provides us with other exactly solvable potentials. Among them, there are trigonometric Scarf and trigonometric Pöschl-Teller potentials which can be prototypes for periodic GWs \cite{phonebook}. We would like to study DME of periodic GWs using them. Another interesting direction is the double copy \cite{Bahjat-Abbas:2017htu} of the Scarf (\ref{Kscarf}) and of the PT (\ref{KPT2}) profiles that we have worked out. We wonder what kind of singular electromagnetic configurations they yield and what their helicities will be. 

So far we studied the geodesic equations of spinless test particles (\ref{geoplane}) based on that the DME conditions of plane waves match the ones in non-relativistic Schroedinger theory. However, there is also supersymmetry in Pauli and Dirac equations \cite{DHoker:1983zea, DHoker:1985xzl, Feher:1989xw}. Thus,  our studies might be extended to trajectories of spinning particles along those arguments.   

As is known, Pöschl-Teller equation frequently occurs in soliton physics. We would like to understand whether there is a relation between the DME of PT profile and the topology of solitons. Likewise, it would be interesting to understand the role of the Schwarzian derivative in \cite{BenAchour:2024ucn} for the Scarf GW. Finally, there is a recent  type of memory called the \emph{frequency memory effect} \cite{Chakraborty:2020uui} in plane GWs. We intend to study such effects in our SUSY framework in the future.

\section*{Acknowledgments}
We are grateful to Krzysztof Andrzejewski, Janos Balog, Peter Horvathy, Cem Yetişmişoğlu and Zurab Silagadze for their comments.
 \vspace{0.2cm}

 \noindent M. E. is supported by İstanbul Bilgi University research fund (BAP) with grant no: 2024.01.009 and by The Scientific and Technological Research Council of Turkey (TÜBİTAK) under grant number 125F021. M. E. is also funded partially by the TÜBİTAK BİDEB 2232-A program under project number 121C067.
M. M. is supported by the Scientific and Technological Research Council of Turkey (TÜBİTAK) under 
grant numbers 122F451 and 123N952.
E. C. is partially funded by the Scientific and Technological Research Council of Turkey (TÜBİTAK) under grant number 125F021.


\vspace{1cm}
\noindent\textbf{Data Availability Statement:}  No data associated in the manuscript.

\pagebreak

\end{document}